\documentclass[conference]{IEEEtran}

\usepackage{float}
\usepackage{amsmath}
\usepackage{tabu}
  \usepackage{caption}
\ifCLASSINFOpdf
  \RequirePackage[pdftex]{graphicx}
  \RequirePackage[absolute]{textpos}
  \RequirePackage{ifthen}
  \RequirePackage[pdftex]{color}
  \RequirePackage{fancyvrb}
  \RequirePackage{amssymb}
  \RequirePackage{multibib}
  \RequirePackage{listings}
  \RequirePackage{booktabs}
  \RequirePackage{amsmath}

\else
  
\fi

\begin{document}
\pagestyle{plain}
\title{Real-time Predictive Maintenance for Wind Turbines Using Big Data Frameworks}

\author{\IEEEauthorblockN{Mikel Canizo\IEEEauthorrefmark{1},
Enrique Onieva\IEEEauthorrefmark{2},
Angel Conde\IEEEauthorrefmark{3}, 
Santiago Charramendieta\IEEEauthorrefmark{4} and
Salvador Trujillo\IEEEauthorrefmark{5}}
\IEEEauthorblockA{\IEEEauthorrefmark{1}\IEEEauthorrefmark{3}\IEEEauthorrefmark{4}\IEEEauthorrefmark{5}IK4-Ikerlan Technology Research Centre\\
Big Data Architectures Area\\
P\textordmasculine. J. M. Arizmendiarrieta, 2. 20500 Arrasate-Mondragón\\
Email: \IEEEauthorrefmark{1}mcanizo@ikerlan.es,
\IEEEauthorrefmark{3}aconde@ikerlan.es,
\IEEEauthorrefmark{4}scharramendieta@ikerlan.es,
\IEEEauthorrefmark{5}strujillo@ikerlan.es}
\IEEEauthorblockA{\IEEEauthorrefmark{2}Deusto Institute of Technology (DeustoTech), University of Deusto\\
Bilbao, Spain \\
Email: enrique.onieva@deusto.es}}

\IEEEoverridecommandlockouts
\IEEEpubid{\makebox[\columnwidth]{978-1-5090-0382-2/16\$31.00 {\textcircled{c}}2017 European Union
		\hfill} \hspace{\columnsep}\makebox[\columnwidth]{ }}
\maketitle

\IEEEpeerreviewmaketitle

\begin{abstract}
		This work presents the evolution of a solution for predictive maintenance to a Big Data environment. The proposed adaptation aims for predicting failures on wind turbines using a data-driven solution deployed in the cloud and which is composed by three main modules. (i) A predictive model generator which generates predictive models for each monitored wind turbine by means of Random Forest algorithm. (ii) A monitoring agent that makes predictions every 10 minutes about failures in wind turbines during the next hour. Finally, (iii) a dashboard where given predictions can be visualized. To implement the solution Apache Spark, Apache Kafka, Apache Mesos and HDFS have been used. Therefore, we have improved the previous work in terms of data process speed, scalability and automation. In addition, we have provided fault-tolerant functionality with a centralized access point from where the status of all the wind turbines of a company localized all over the world can be monitored, reducing O\&M costs.

		\begin{IEEEkeywords}
			Big Data architectures, Machine learning, Cloud computing, Wind power, Industry 4.0.
		\end{IEEEkeywords}
		
\end{abstract}

\section{Introduction}
The growth of renewable energy remains very fast, especially in the wind industry, where the increase has been exponential in recent years. This aspect can be reflected on and considered upon the wind energy power that was deployed from 1997 to 2014. This trend is expected to continue in the future \cite{kaldellis_shifting_2013}.

 In fact, global wind energy installations totalled 433 GW as of the end of 2015, and the industry is set to grow by another $\sim$60 GW in 2016 \cite{council_2016}. Moreover, the Energy Roadmap 2050 proclaims an aim that wind energy supplies of between 31.6\% and 48.7\% of Europe's electricity will be achieved \cite{Garcia_marquez_2012}. This increase has had a great impact on the operation and maintenance (O\&M) costs, where within the energy generation costs, O\&M costs can reach up to 32\% \cite{kaldellis_shifting_2013}.
 
 In the current situation, it is very important that for both medium and large-sized industrial companies, that there is an implementation of predictive maintenance strategies for increasing the lifecycle of their wind systems. This scenario will increase the lifecycle of the company's systems, improving their availability and reliability, which will then directly affect the productivity \cite{hashemian_state_art_2011}. In addition, such strategies will reduce O\&M costs. However, the growth of data that has to be analyzed in order to do a predictive maintenance is proportional to the growth of the wind industry. Currently, the daily data volumes that are generated by the wind turbines are too large to be processed with traditional technology \cite{provost_2013}. The management of these volumes of data requires a system capable of incorporating the entire technology stack: the extraction-transformation-load (ETL), the data filtering, the data production, the statistical models and the data mining. These all need to be embedded into a platform capable of managing multiples tables of millions of rows, 10-100 times faster than traditional technology. The key is to get the correct information in the right format and at the right moment \cite{Faiz_2009}. At this point, Big Data frameworks come in for analyzing the data more efficiently so decision making processes can be improved. Besides, it has already been shown that there is an increment of 3\% on a company's productivity by using Big Data frameworks \cite{provost_2013}.
 
 This work presents a Big Data approach to adapt a predictive maintenance method for wind turbines in order to be executed in a scalable cloud computing environment. This way, it can be deployed easily on a cloud such as Amazon EMR\footnote{https://aws.amazon.com/emr/?nc1=h\_ls} or Microsoft HDInsight\footnote{https://azure.microsoft.com/en-us/services/hdinsight/} and scale horizontally as the data to be processed increases its volume. The article shows how the original solution was improved in almost all aspects. Besides improving the method in terms of speed, scalability, automation and reliability, we obtained better results as it has higher overall accuracy and sensitivity rate \cite{kusiak_prediction_2011}.
 
 This article is structured as follows: Section \ref{section-state-of-art} presents the State of the Art. Section \ref{section-methodology} describes the usages of Big Data architecture and followed methods. Section \ref{section-experimentation} analyzes the obtained results and the applied studies. Section \ref{section-conclusions} exposes the conclusions and the future work.
 
\section{State of Art} \label{section-state-of-art}
Predictive maintenance, sometimes categorized as "\textit{online monitoring}", has a large history. Visual inspection, while being the oldest method, is still the most used one. These have now evolved into automated methods by using processing techniques that are based on advanced pattern recognition signals \cite{ hashemian_state_art_2011}.

Over recent years, about 30\% \cite{hashemian_state_art_2011} of the industrial equipment does not benefit from predictive maintenance techniques. Instead, periodic maintenance is used in order to detect any anomalies or malfunctions in the components of their systems. Such maintenance is usually done visually and physically placed on the machine. In order to clearly discern between a periodical and predictive maintenance, it has been shown that no problems were found in 70\% \cite{estado_arte_hale} of the periodic revisions, while the percentages have reached up to 90\% \cite{hashemian_advanced_1998} when using predictive maintenance techniques. This suggests that the latter method can increase the maintenance efficiency, and therefore, it can reduce the amount of failures in industrial systems.

Although the maintenance based on periodic revisions is the most extended and used method, these techniques are being increasingly classified as constituting defective and unreliable methods \cite{mobley_introduction_2002}. After conducting a study with identical systems that were tested under identical conditions \cite{hashemian_state_art_2011}, it has been shown that the time until a failure occurs in the system is very different from one system to another. The maintenance that is based upon periodic revisions is thus ineffective, because it is very difficult to know when a component of an industrial process is going to fail based on a fixed period of time.

The evolution of technology has made predictive maintenance techniques evolve too. The use of wireless sensors and the posterior use of Supervisory Control and Data Acquisition (SCADA) systems have provided companies with new ways of collecting information about the performances of their industrial machines. With these systems, more data can be gathered in an easier manner. Therefore, the volumes of available data are larger. One problem that arises is how to know which data is relevant and which is not. As a result, the challenge is to know how to obtain valuable information from such data so as to support the decision making processes. For that kind of a solution, data mining algorithms began to be used in order to extract such information from historical data. Data mining algorithms allow to see the behavior of industrial equipment over time and to predict future failures, based on the extracted information. Over the years, predictive models have been generated by using different techniques, which have been classified into three main areas: model based, data-driven based, and case based models \cite{cheng_case-based_2015}. Nowadays, data-driven based models are the most used ones \cite{hashemian_state_art_2011}.

In the field of wind energy, where wind turbines are becoming larger and more powerful, having a good failure detection system has become indispensable. In recent years, there have been several studies in which new methods have been proposed for performing a predictive maintenance on wind turbines using data mining algorithms. Focusing on previous classified techniques, within predictive models based on models, several approaches have been identified like linear models combined with artificial neural networks \cite{cross_model-based_2013}, the use of Model Predictive Control (MPC) method \cite{vieira_failure_2013} or Extreme Learning Machine (ELM) algorithms \cite{qian_condition_2015}. Within data-driven techniques, there are also some remarkable researches like an anticipatory control based on MPC approach using time series model \cite{kusiak_anticipatory_2009}, a novel predictive maintenance method based on the Random Forest algorithm \cite{p._verma_performance_2012}, a three phase based method: offline training process, online monitoring phase and online diagnosis phase \cite{kruger_data-driven_2013}, a new method using Artificial Neural Networks (ANN) \cite{bangalore_approach_2013} or one that uses fuzzy models \cite{simani_fault_2014}. Within case based techniques, there is a study that proposes a novel method which combines an Adaptative Neuro-Fuzzy Inference System (ANFIS) with a Big Data paradigm \cite{cheng_case-based_2015}.

In the last years, Big Data analytic approaches are becoming more popular in a wide variety of industries and purposes such for medicine \cite{bates_big_2014}\cite{westra_nursing_2015}, social networks recommendations \cite{jiang_big_2016}, logistic \cite{ben_ayed_big_2015}, structural health monitoring \cite{cai_bigdata-analytics_2016} \cite{alampalli_big_2016}, business strategy development \cite{zhong_apply_2015} or power consumption in manufacturing \cite{shin_predictive_2014}. For the wind energy industry, some studies have been also identified where a business intelligence approach is presented for failure prognosis \cite{helsen_long-term_2016} or the requirements for a Big Data approach are analyzed \cite{ nabati_bigdata-analytics_2016}.
\section{Methodology} \label{section-methodology}
This section presents the methodology that has been followed within this development. Section \ref{subsection-motivation} relates to the motivation of this work. Section \ref{subsection-problem} describes the problems that arise. Section \ref{subsection-data} exposes the data that has been used. Section \ref{subsection-architecture} and Section \ref{subsection-method} describe, respectively, the architecture designed and the method followed to solve the presented problem.

	\subsection{Motivation} \label{subsection-motivation}
	As described in Section \ref{section-state-of-art}, there are many approaches which present novel methods to perform a predictive maintenance for wind turbines. Although there are some researches using Big Data approaches, we could not find nothing in the literature using the latest technologies like the ones described in this article, to apply a complete solution for the predictive maintenance of wind turbines. Thus, our motivation is to contribute with an adoption of an existing reliable predictive maintenance method \cite{ p._verma_performance_2012} to a Big Data environment. Therefore, the application can be executed in the cloud and it can easily scale as much as needed.

	\subsection{Problem} \label{subsection-problem}
	With the growth of the wind energy power, the companies of this industry have more and more wind turbines. Moreover, wind turbines are bigger and more powerful. Thus, the companies need to add more resources for maintenance purposes, increasing O\&M costs. A predictive maintenance system based on traditional technologies requires having one system in each wind farm since they do not use neither private nor public cloud where centralize all the data, so that the company can analyze them from a unique location. In addition, the company needs to hire qualified people to manage each predictive maintenance system.
	
	However, having all the data produced by the wind turbines placed in a central system involves a huge computational cost to being capable of processing all the information fast enough to notify a future failure. Moreover, the notification has to arrive with enough time allowed to repair it before the component breaks down.
	
	Other problem of traditional technologies is the scalability. If the volume of data to be processed increases, it is very hard to add more computational resources to handle it. In this situation, the scalability is usually provided by buying a more powerful hardware (vertical scalability), which is very expensive and it has a limited scalability. This limit is set by the current technology.
	
	The work we have selected to adapt to a Big data environment, has these all problems because it was developed with traditional technologies. To avoid them, we have developed an application using cloud computing and Big Data frameworks to provide the method with the ability to scale in an easier and cheaper way (horizontal scalability) and to process the data of hundreds of thousands of wind turbines in a scalable way. Finally, these technologies let a company perform the predictive maintenance of all its wind turbines located all around the world from a central system in a fault-tolerant manner. This will mean a considerable cost reduction in O\&M services.

	\subsection{Data} \label{subsection-data}
	The data used in this work is composed by data gathered from wind turbines during a period of two years. Data is divided in two types: status data (alarms activations and deactivations) and operational data, which represent the performance of the wind turbines. There are 448 different alarm types and 104 parameters concerning the operational data.
	
	New data is received every 10 minutes, so the values of received data represent the mean values of such period. Therefore, a wind turbine that works all the day will send 144 (6x24) rows of data per day. This work uses data from a wind farm of 17 wind turbines, so 2.448 new rows of data are obtained every day for a total of 1.787.040 in two years, in the best case. The number of daily rows of data may not seem very high, but this application is designed for scaling to a hundreds or thousands of wind turbines, which means a significant volume of data to be processed.

	\subsection{Architecture} \label{subsection-architecture}
\begin{figure}[!h]
		\centering
		\includegraphics[width=.4\textwidth]{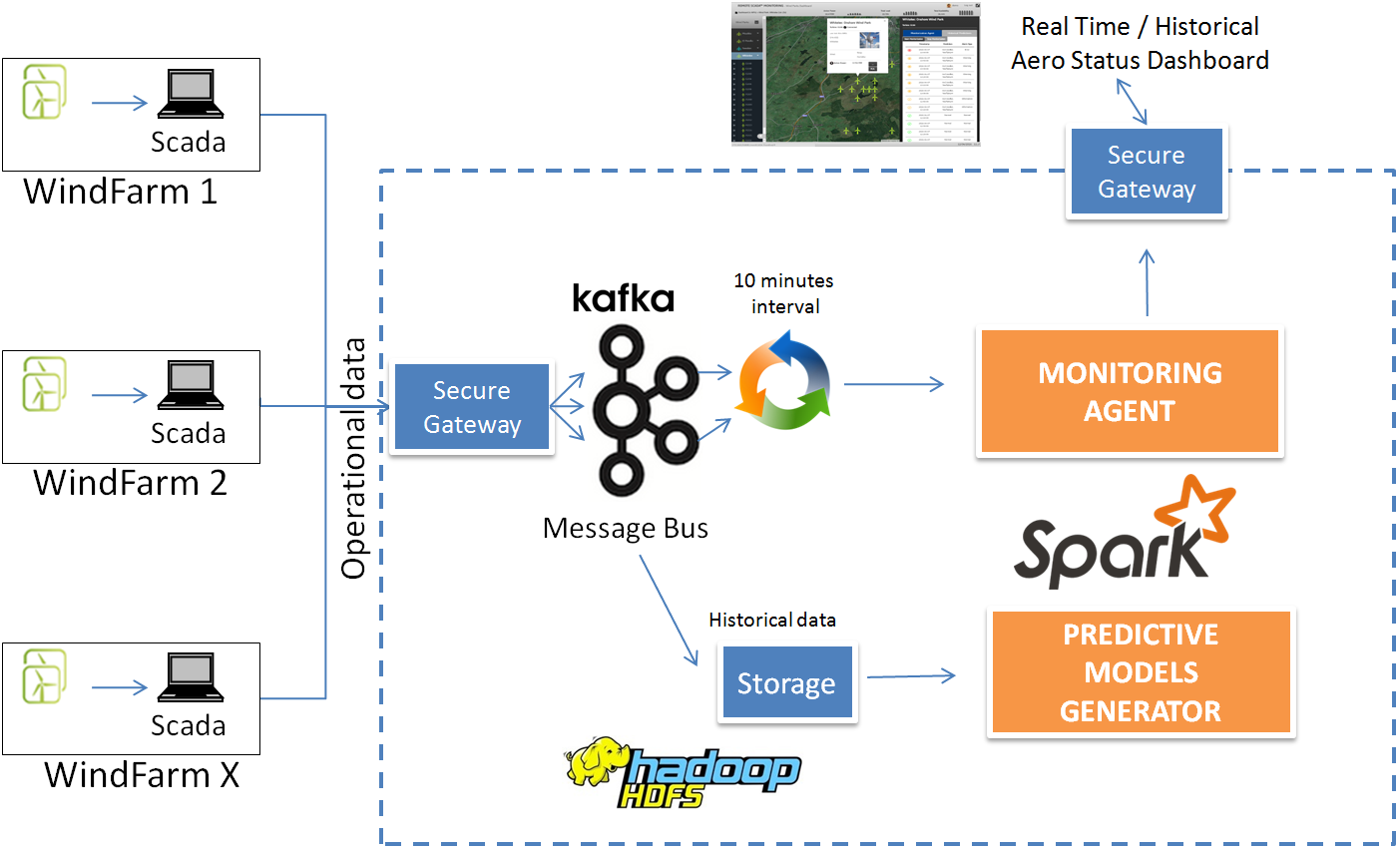}
		\caption{Application architecture}
		\label{fig:Arquitectura}
	\end{figure}

	To design the Big Data architecture for this application, data acquisition, data persistence and data processing aspects had to be taken into account. This was in order to make the predictive maintenance to be remotely distributed, efficient, scalable, and fault tolerant. Figure \ref{fig:Arquitectura} shows the architecture of the application and the used technologies.
	
	For the data persistence, the Hadoop Data File System (HDFS)\footnote{http://hadoop.apache.org/docs/r1.2.1/hdfs\_design.html} is used. HDFS is a distributed file system that is designed to run on commodity hardware. HDFS is fault-tolerant and and can scale horizontally. Thus, HDFS is suitable for this application, because it has to persist and load large volumes of data that are obtained from the wind turbines, in a distributed manner, even if any errors occurs. 	

	HDFS has master/slave architecture. It uses a single NameNode which works as a master server and some DataNodes which work as slaves. NameNode manages the file system namespace and regulates access to the files by clients. DataNodes manage the storage that is attached to the nodes as they run. The data is formulated into files that are split into one or more blocks. These blocks are stored in one or more DataNodes. In order to formulate the data that is gathered from the wind turbines, each turbine has two files: one for the status data and the other one for the operational data.
	
	For the data acquisition, Apache Kafka\footnote{https://kafka.apache.org/} is used. Apache Kafka is a distributed streaming platform which works as a messaging system. It uses a public/subscriber distributed message system that classifies the messages into topics. A topic is like a channel or a pipeline where messages are published. The topics on Kafka are always multi-subscriber. That is, a topic can have zero, one, or many consumers, subscribed to the data written on it. However, this application only uses one topic for each wind turbine, so there is only one publisher and one consumer per topic. On the one hand, the wind turbines publish data to the corresponding topics. On the other hand, the monitoring agent has a consumer for each wind turbine that it is monitoring. This way, the data for each wind turbine is well organized.
	
	Moreover, Kafka is a fault-tolerant system. If any error occurs, published messages are still available---whether or not they have been consumed---when the application or Kafka is restarted. Thus, Kafka can be guaranteed not to miss any message that is published from the wind turbines. Hence, the consumers will consume the published messages at least once. So the application can make a prediction for each published operational data, even if an error has occurred.
	
	For data processing purposes, Apache Spark\footnote{http://spark.apache.org/} is used. Spark is an in memory fast and general engine for large-scale data processing. It can process data 100 times faster than traditional technologies, according to the official documentation. In addition, Spark is a scalable system that can scale to hundreds or thousands of machines in a fault-tolerant manner. These characteristics make Apache Spark very suitable to process and analyze the large volumes of data that are gathered from the wind turbines. The applications can be developed in Scala, Java, Python and R, but Java is used for this application. Spark is composed of five modules: Spark Core, Spark Streaming, Spark SQL, MLlib and GraphX. So Spark provides the requisites required for developing the proposed approach that is embedded into one framework.
		
	This application uses two data processing types: offline processing in order to generate the predictive models and online processing to make predictions in real time. Spark Core is used for the offline mode, while Spark Streaming is employed for the online mode. For querying the loaded data from the HDFS, Spark SQL is used. In addition, Spark's machine learning library (MLlib) is used for data mining purposes.

	For the cluster management purpose, Apache Mesos\footnote{http://mesos.apache.org/} and Apache Zookeeper\footnote{https://zookeeper.apache.org/} are used. Apache Mesos abstracts CPU, memory, storage, and other computational resources away from the machines (physical or virtual), enabling fault-tolerant and elastic distributed systems. Thus, it provides for a way to unify all of the used frameworks as a single pool of resources. Apache Mesos manages the resources of the clusters that the frameworks could use. Apache Spark has two modes of execution in Apache Mesos: a coarse-grained one and a fine-grained one. The first mode allocates the fixed resources and they are reserved only for it until the application finishes. The second mode manages the resources dynamically, so that when the application requires more resources, Mesos assigns them to it until there is a specified maximum of resources. Otherwise, if the application does not need so many resources, it will leave them so that other applications can use them. For this development, the fine-grained mode was selected, because the application runs in a private cloud, the same as other applications. Hence, all of the applications have to live with each other, sharing the fixed resources.
	
\begin{figure}[!h]
		\centering
		\includegraphics[width=.4\textwidth]{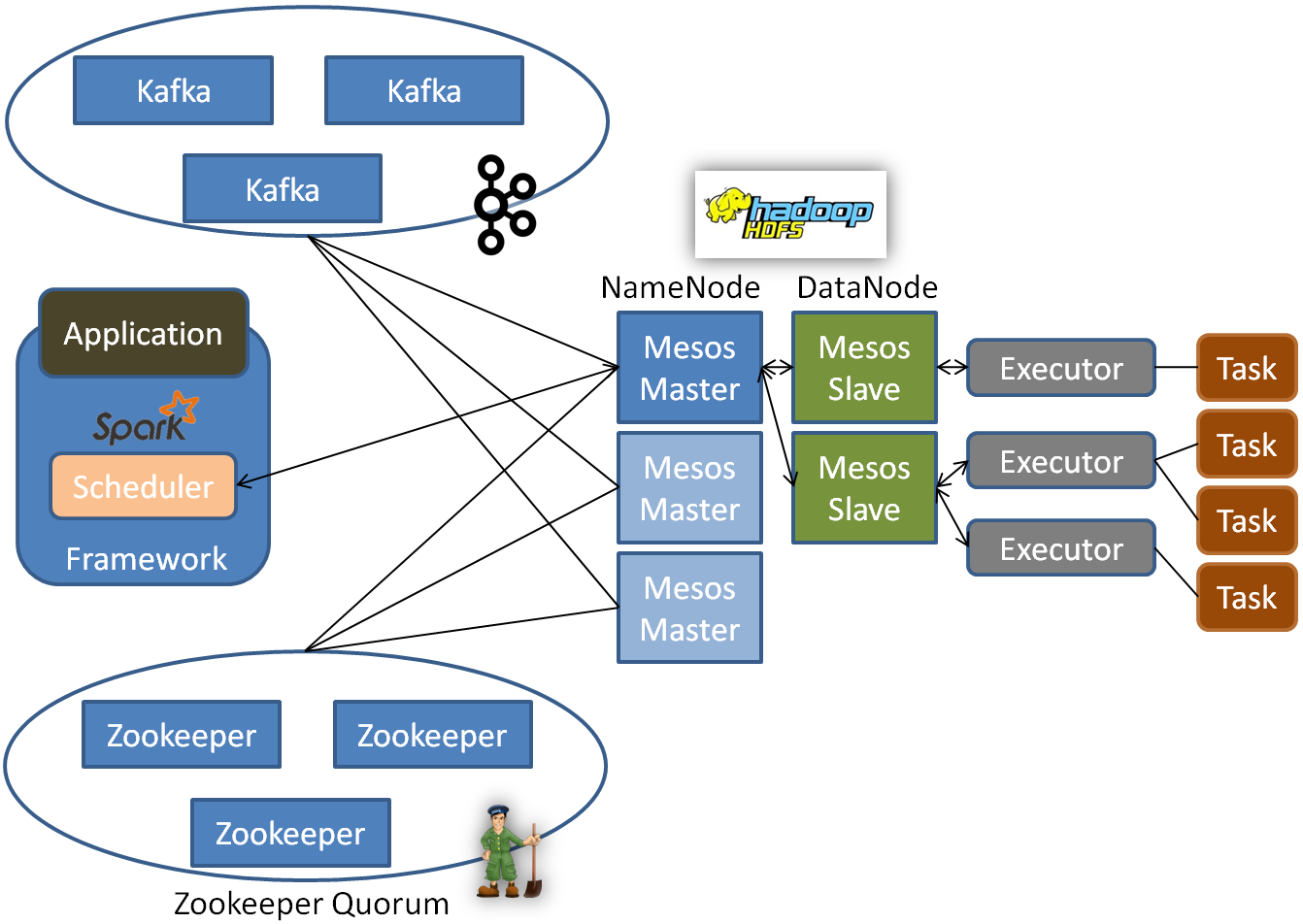}
		\caption{Cluster architecture}
		\label{fig:cluster-architecture}
	\end{figure}

	Apache Zookeeper is a centralized service for maintaining the configuration information and the naming for providing distributed synchronization and group services. In this particular case, it is used for load balancing and for the Mesos master node selection purpose. As is shown in Figure \ref{fig:cluster-architecture}, there are three Apache Mesos master nodes and two slave nodes. Only one master node runs at a time, so the other two nodes are in a standby mode waiting to start running if the principal node fails. In that case, Zookeeper will select the new master node. For that selection process, there are three Zookeeper nodes to make the decision in a quorum. It could work with only one node, but three nodes are the minimum number of nodes for performing a quorum. Besides, having three nodes provides for load balancing and for fault-tolerant features.
	
	Referring to the execution of the application within the cluster, the master node contains the main part of the application and the slave nodes have executors, which execute the required jobs or tasks. So, when a function is required to be executed in the master node, it is divided into one or more tasks. These tasks are sent to the slave nodes to be executed by the executors and the result of them, converge to the principal node when needed.

	This cluster is installed in a private cloud which has 32 cores and 128GB of RAM. Six of these cores are assigned to Kafka and Zookeeper, three for each one. Mesos and HDFS share 5 cores and the rest are reserved for the applications. The proposed Spark application uses at most, 6 cores and 30GB of RAM. Note that HDFS's NameNode is installed in the same node as the Mesos master nodes and that HDFS's DataNodes are installed in the same nodes as the Mesos Slaves. Hence, the data that the application has to interact with is stored in the slave nodes. As tasks are executed in the slave nodes, having data already placed in those nodes improves the efficiency of the application, because the data may have not to be sent through the network.

	\subsection{Method} \label{subsection-method}
	To better understand this section, the reader is urged to read this article \cite{kusiak_prediction_2011}. It describes the predictive maintenance method we have adapted and modified to deploy it in a cloud by means of Big Data frameworks.
	
	Our method is divided into three parts: a predictive models generator which generates the predictive models that are based upon the historical data from the wind turbines; a monitoring agent which makes the predictions when the data is received from the wind turbines (every 10 minutes); and a front-end where the wind turbines are able to be visualized geographically located in a map, as well as their notifications about the predictions and the status information. While the first two parts follow the original method, the last one is a new functionality added to build a complete application.
	
\begin{figure}[!h]
		\centering
		\includegraphics[width=.4\textwidth]{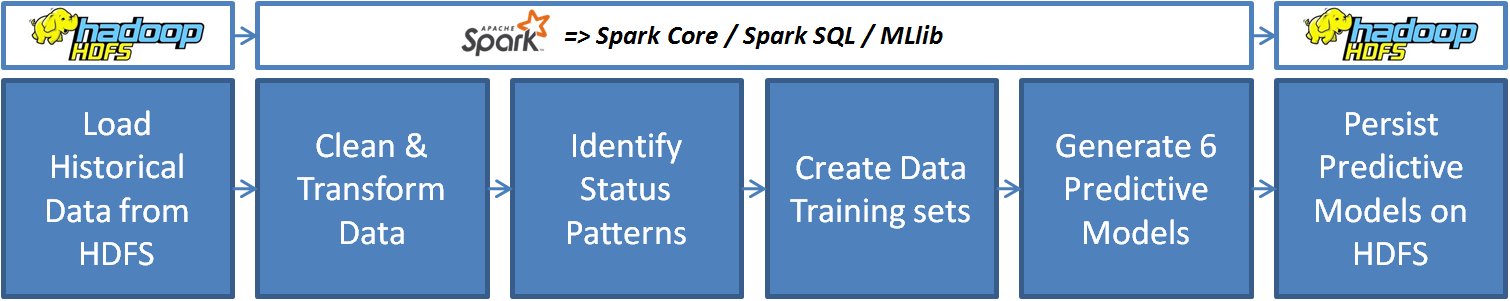}
		\caption{Predictive Model Generator diagram}
		\label{fig:model-generator-diagram}
	\end{figure}

	As shown in Figure \ref{fig:model-generator-diagram}, the predictive model generator obtains the historical alarms data of each wind turbine from the HDFS and it performs some ETL processes in order to remove useless data and to give them an adequate format. An association rules algorithm is then applied to identify the critical status patterns that involve a fault on a wind turbine. For this purpose, only status data is used.
	
	Then, the base data training set is constructed combining identified status patterns and operational data. Following the original method, six data training sets are made and Random Forest algorithm is applied to them to generate the required six predictive models. An experimentation was done to identify which were the best parameters to apply the algorithm to this use case. Finally, each model is persisted in the HDFS. This process was done in an automated, distributed and concurrent way.
	
\begin{figure}[!h]
		\centering
		\includegraphics[width=.4\textwidth]{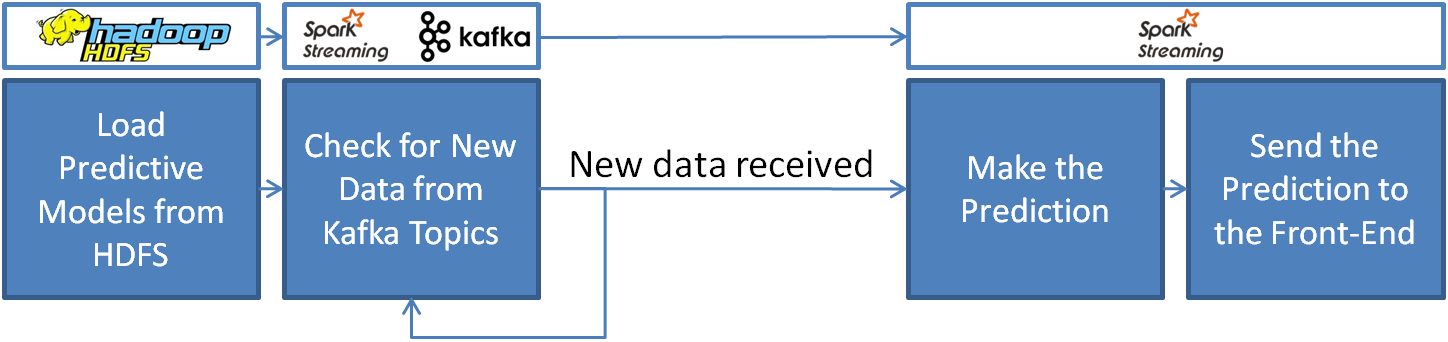}
		\caption{Monitoring Agent diagram}
		\label{fig:monitoring-agent-diagram}
	\end{figure}

	In the second part, the monitoring agent works 24*7*365. As shown in Figure \ref{fig:monitoring-agent-diagram}, at first, it loads all predictive models in memory so that the agent can start to monitor the state of the wind turbines. It means that the agent can then process the data received from Apache Kafka's topics. When new data is received, a prediction is made by means of the corresponding predictive models. Note that the wind turbines send operational data every 10 minutes and that the prediction is always done to forecast what is going to happen one hour later. Finally, each prediction is sent via websockets to the front-end for its visualization. As the agent has to keep working continuously, it is designed to automatically recover its normal activity if something wrong happens.
	
\begin{figure}[!h]
		\centering
		\includegraphics[width=2.7in]{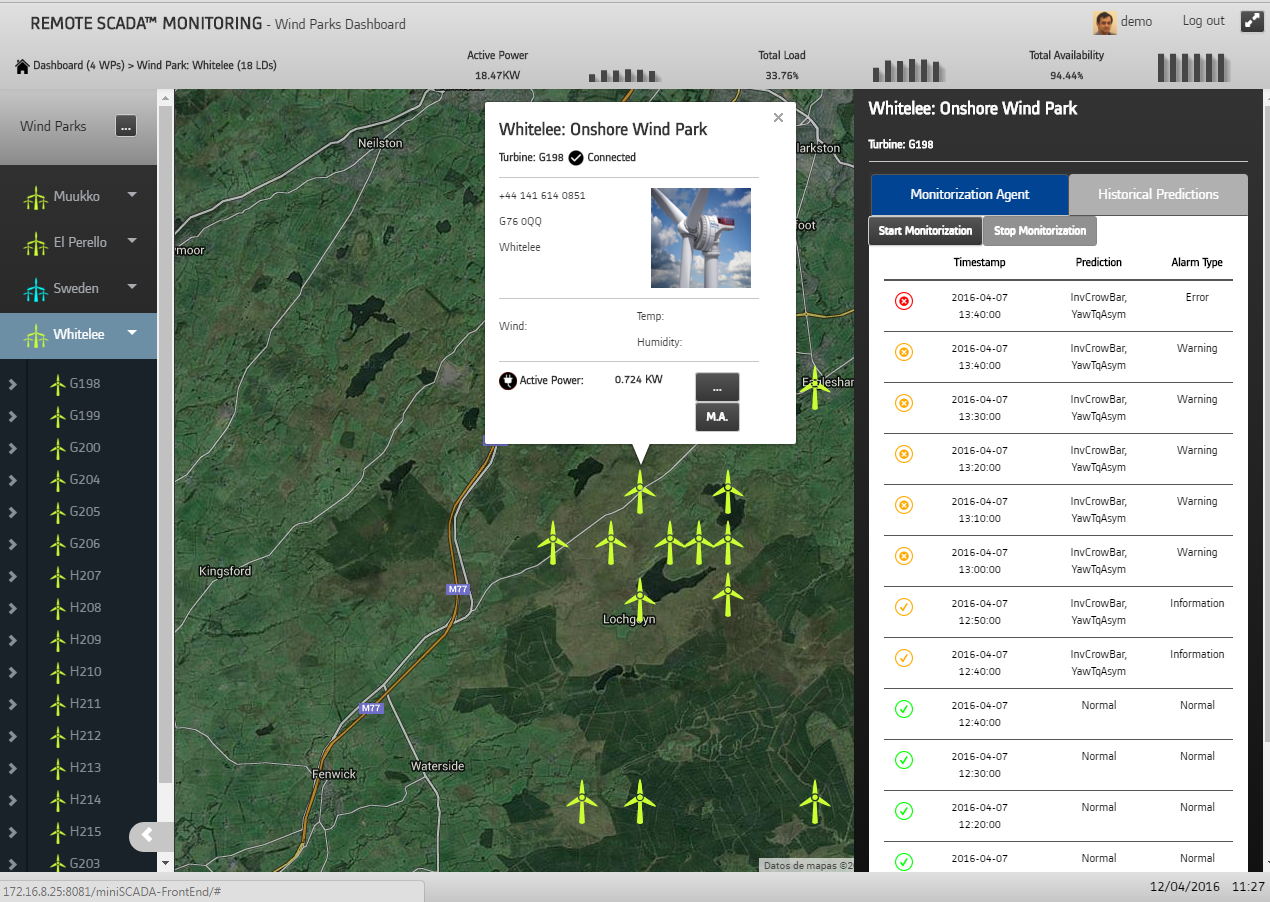}
		\caption{Front-end}
		\label{fig:FrontEnd}
	\end{figure}

	In the third part, the front-end that was developed shows user's wind farms geographically located on a map. Each wind farm has its own configuration and its own wind turbines. So, the state of the wind farms and its wind turbines can be visualized in real time, as well as the notifications about the performed predictions. In Figure \ref{fig:FrontEnd}, a front-end's screen-shot is shown.

\section{Experimentation} \label{section-experimentation}
In this section, the results of the experimentation are specifically described. Section \ref{subsection-parameters_selection} describes the performed analyzes for selecting the most relevant parameters. Section \ref{subsection-random_forest} exposes how the Random Forest's parameter values were selected. Section \ref{subsection-results} shows the accuracy of the predictive models. Section \ref{subsection-discussion} discusses the obtained results.

	\subsection{Parameters Selection} \label{subsection-parameters_selection}
\begin{figure}[!h]
		\centering
		\includegraphics[width=.31\textwidth]{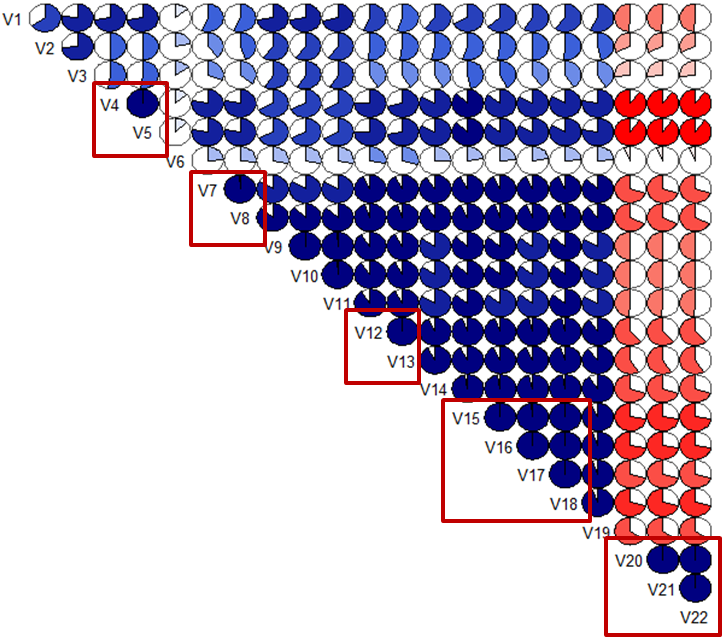}
		\caption{Pearson's coefficient correlation}
		\label{fig:Correlacion}
	\end{figure}

	As described in the previous sections, the operational data has 104 parameters, but not all of them are relevant for the predictive models. Thus, a Principal Component Analysis (PCA) was instigated in order to identify the most relevant ones. This way, the list of parameters was reduced to 22, because these variables represented the 99\% of accumulated covariance. To verify these results, Pearson’s correlation coefficient analysis was performed. The obtained results showed that there were some groups with similar parameters, as is shown in Figure \ref {fig:Correlacion}. So only one of each group was selected and the other ones were discarded. In such a manner, the list was reduced to 14 parameters.
	
	With having to process fewer variables, the required time to generate the predictive models decreased significantly. This fact led us to improve the accuracy of the predictive models because only the relevant parameters were used.

	\subsection{Selection of Random Forest parameters value} \label{subsection-random_forest}
\begin{figure}[!h]
		\centering
		\includegraphics[width=1.75in]{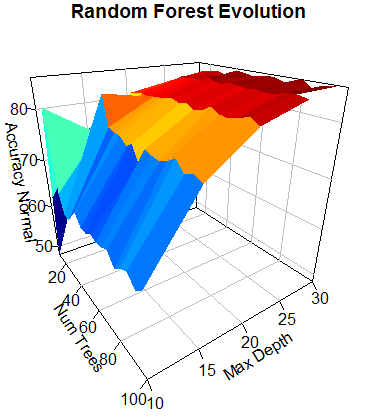}
		\caption{Accuracy of the normal output vs. \textit{numTrees} and \textit{maxDepth}}
		\label{fig:grafico-patron-normal}
	\end{figure}

\begin{figure}[!h]
		\centering
		\includegraphics[width=1.75in]{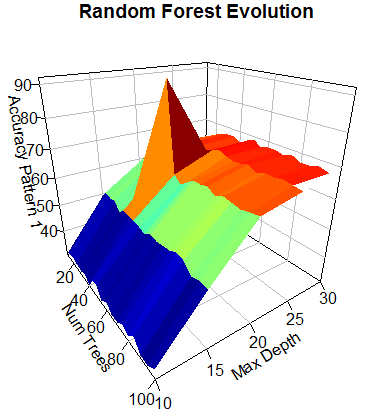}
		\caption{Accuracy of the status pattern 1 vs. \textit{numTrees} and \textit{maxDepth}}
		\label{fig:grafico-patron-1}
	\end{figure}

\begin{figure}[!h]
		\centering
		\includegraphics[width=1.75in]{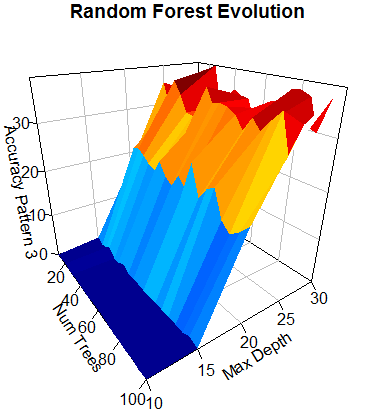}
		\caption{Accuracy of the status pattern 3 vs. \textit{numTrees} and \textit{maxDepth}}
		\label{fig:grafico-patron-3}
	\end{figure}

\begin{figure}[!h]
		\centering
		\includegraphics[width=1.75in]{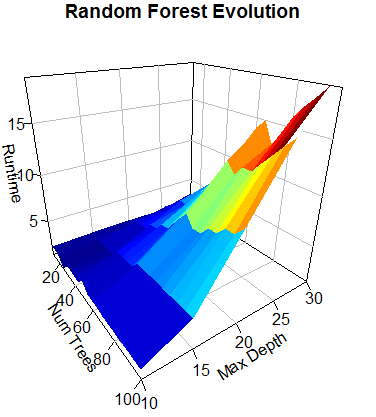}
		\caption{Computational cost vs. \textit{numTrees} and \textit{maxDepth}}
		\label{fig:grafico-coste-computacional}
	\end{figure}

	The Random Forest algorithm was used in order to generate the predictive models. $N_{trees}=40$ and $Max_{depth}=25$ were selected as the parameter values. These values were obtained from a performed analysis where all of the predictive models were tested by using the following values: $N_{trees}=$\{5,10,15,20,...,100\} and $Max_{depth}=$\{5,10,15,...,30\}. The obtained results showed that the parameter which had the most impact on the model's precision was the depth of the trees. Figures \ref{fig:grafico-patron-normal}, \ref{fig:grafico-patron-1} and \ref{fig:grafico-patron-3} graphically show the results that were obtained in terms of accuracy, as the values of the previous variables increased. As is shown, the increment in the number of trees did not very much affect the precision, while the increment in the depth of the trees had a significant impact on the improvements of the precision of the predictive models. Figure \ref{fig:grafico-coste-computacional} shows the behavior of the computational cost as the values of both parameters were increased. As is shown, the computational cost may reach up to 20 minutes per each predictive model.

	\subsection{Obtained Results} \label{subsection-results}
	The accuracy of the predictive models was measured by the overall success, the success rates of errors (status pattern), no errors (normal), and the success rates of individual classes. In addition, the sensitivity and the specificity of each model is measured. The sensitivity measures how good a model is at detecting positives while the specificity measures how good a model is at avoiding false alarms. In the following tables, the results of three wind turbines are shown. Note that each wind turbine had its own identified status patterns. Each status pattern corresponded to a specific class. In Tables \ref{tab:pruebas-aero-5}, \ref{tab:pruebas-aero-6} and \ref{tab:pruebas-aero-7}, the model's number represents the time-stamp of the predictive models that went from t+10 to t+60 (1-6). The percentages that appear within the number of the classes are the percentages of the instances for each class within the dataset.

	\begin{itemize}
		\item \textbf{Number of classes:} 5 (Table \ref{tab:pruebas-aero-5})
		
		\item \textbf{Classes:}
		
		\begin{itemize}
			\item \textbf{Class 1:} \textit{InvCH0Loss, WLFRTActive}
			
			\item \textbf{Class 2:} \textit{InvCH0Loss, WLFRTActive, GOverSpMax}
			
			\item \textbf{Class 3:} \textit{GOverSpMax ,WLFRTActive}
			
			\item \textbf{Class 4:} \textit{Normal}
			
			\item \textbf{Class 5:} \textit{GOverSpMax, InvCH0Loss}
		\end{itemize}
	\end{itemize}

	\begin{table*}[h]
		\centering
		\caption{Accuracy of predictive models of wind turbine 5}
		\begin{center}
			\begin{tabu} to 0.95\textwidth{X[c] | X[c] X[c] X[c] X[c] X[c] | X[c] X[c] | X[c] | X[c] | X[c]}
				\toprule
				Accuracy type & \multicolumn{5}{c}{Accuracy individualized by class} & \multicolumn{2}{|c|}{Accuracy error/no error} & Accuracy & Sensitivity & Specificity \\
				\midrule
				Models \textbackslash{} Class & Class 1 (11.74\%) & Class 2 (50.01\%) & Class 3 (0.75\%) & Class 4 (37.24\%) & Class 5 (0.26\%) & Error (62.76\%) & No error (37.24\%) & Global & Global & Global \\
				\midrule
				Model 1 & 62.60\% & 92.56\% & 18.52\% & 82.35\% & 6.12\% & 85.52\% & \multicolumn{1}{c|}{82.35\%} & 84.34\% & 91.54\% & 74.11\% \\
				Model 2 & 64.50\% & 92.76\% & 9.96\% & 81.93\% & 9.09\% & 86.08\% & \multicolumn{1}{c|}{81.93\%} & 84.54\% & 92.10\% & 73.80\% \\
				Model 3 & 60.69\% & 92.59\% & 12.88\% & 82.13\% & 4.49\% & 85.63\% & \multicolumn{1}{c|}{82.63\%} & 84.32\% & 91.61\% & 74.04\% \\
				Model 4 & 65.85\% & 92.40\% & 15.87\% & 82.32\% & 5.95\% & 86.18\% & \multicolumn{1}{c|}{82.32\%} & 84.74\% & 92.15\% & 74.24\% \\
				Model 5 & 63.43\% & 92.51\% & 24.27\% & 82.01\% & 4.30\% & 85.91\% & \multicolumn{1}{c|}{82.01\%} & 84.45\% & 92.15\% & 73.69\% \\
				Model 6 & 64.95\% & 92.47\% & 16.12\% & 82.58\% & 8.82\% & 85.90\% & \multicolumn{1}{c|}{82.58\%} & 84.67\% & 91.98\% & 74.30\% \\
				\bottomrule
			\end{tabu}%
		\end{center}
		\label{tab:pruebas-aero-5}%
	\end{table*}%

	\begin{itemize}
		\item \textbf{Number of classes:} 2 (Table \ref{tab:pruebas-aero-6})
		
		\item \textbf{Classes:}
		
		\begin{itemize}
			\item \textbf{Class 1:} \textit{Normal}
			
			\item \textbf{Class 2:} \textit{YawTqAsym, WLFRTActive}
			
		\end{itemize}
	\end{itemize}

	\begin{table*}[h]
		\centering
		\caption{Accuracy of predictive models of wind turbine 6}
		\begin{tabu} to 0.95\textwidth{X[c] | X[c] X[c] | X[c] X[c] | X[c] | X[c] | X[c]}
			\toprule
			Accuracy type & \multicolumn{2}{c}{Accuracy individualized by class} & \multicolumn{2}{|c|}{Accuracy error/no error} & Accuracy & Sensitivity & Specificity \\
			\midrule
			Models \textbackslash{} Class & Class 1 (47.44\%) & Class 2 (52.56\%) & Error (52.56\%) & No error (47.44\%) & Global & Global & Global \\
			\midrule
			Model 1 & 77.94\% & \multicolumn{1}{c|}{89.04\%} & 89.04\% & 77.94\% & 83.78\% & 89.04\% & 77.94\% \\
			Model 2 & 78.15\% & \multicolumn{1}{c|}{89.44\%} & 89.44\% & 78.15\% & 84.07\% & 89.44\% & 78.15\% \\
			Model 3 & 77.14\% & \multicolumn{1}{c|}{89.58\%} & 89.58\% & 77.14\% & 83.68\% & 89.58\% & 77.14\% \\
			Model 4 & 77.44\% & \multicolumn{1}{c|}{89.18\%} & 89.18\% & 77.44\% & 83.59\% & 89.18\% & 77.44\% \\
			Model 5 & 77.49\% & \multicolumn{1}{c|}{89.43\%} & 89.43\% & 77.49\% & 83.77\% & 89.43\% & 77.49\% \\
			Model 6 & 77.29\% & \multicolumn{1}{c|}{89.20\%} & 89.20\% & 77.29\% & 83.56\% & 89.20\% & 77.29\% \\
			\bottomrule
		\end{tabu}%
		\label{tab:pruebas-aero-6}%
	\end{table*}%

	\begin{itemize}
		\item \textbf{Number of classes:} 3 (Table \ref{tab:pruebas-aero-7})
		
		\item \textbf{Classes:}
		
		\begin{itemize}
			\item \textbf{Class 1:} \textit{WLFRTActive, YawTqAsym, YawBrBlock}
			
			\item \textbf{Class 2:} \textit{Normal}
			
			\item \textbf{Class 3:} \textit{YawBrBlock, WLFRTActive}
		\end{itemize}
	\end{itemize}
	
	\begin{table*}[h]
		\centering
		\caption{Accuracy of predictive models of wind turbine 7}
		\begin{tabu} to .95\textwidth{X[c] | X[c] X[c] X[c] | X[c] X[c] | X[c] | X[c] | X[c]}
			\toprule
			Accuracy type & \multicolumn{3}{c}{Accuracy individualized by class} & \multicolumn{2}{|c|}{Accuracy error/no error} & Accuracy & Sensitivity & Specificity \\
			\midrule
			Models \textbackslash{} Class & Class 1 (51.54\%) & Class 2 (32.84\%) & Class 3 (15.62\%) & Error \linebreak(67.16\%) & No error (32.84\%) & Global & Global & Global \\
			\midrule
			Model 1 & 96.30\% & 80.17\% & 42.39\% & 83.38\% & 80.17\% & 82.62\% & 97.56\% & 69.59\% \\
			Model 2 & 96.30\% & 80.93\% & 41.59\% & 83.58\% & 80.93\% & 82.66\% & 97.68\% & 69.43\% \\
			Model 3 & 96.29\% & 80.32\% & 42.73\% & 83.81\% & 80.32\% & 81.96\% & 97.62\% & 70.25\% \\
			Model 4 & 96.26\% & 81.08\% & 41.85\% & 83.41\% & 81.08\% & 82.61\% & 97.79\% & 69.39\% \\
			Model 5 & 96.22\% & 80.85\% & 42.69\% & 83.46\% & 80.85\% & 81.87\% & 97.53\% & 70.00\% \\
			Model 6 & 96.47\% & 80.95\% & 42.00\% & 83.53\% & 80.95\% & 82.66\% & 97.70\% & 69.60\% \\
			\bottomrule
		\end{tabu}%
		\label{tab:pruebas-aero-7}%
	\end{table*}%

	\subsection{Results Discussion} \label{subsection-discussion}
	Referring to the overall success, the accuracy of most of the predictive models obtained a success rate of 81\%-85\%, which was higher than the adopted method, since it obtained an overall accuracy of 76\% \cite{kusiak_prediction_2011}. This means a gain of 5\%-9\%. However, the precision of the predictive models for 2 of the 17 wind turbines decreased to 70\%-75\%. When analyzing this fact, we realized that the predictive models of both of these two wind turbines had more of a number of classes than the others, specifically five and six. Therefore, we have concluded with a hypothesis that the more classes there are, the worse the accuracy will be for the predictive models.	
	
	Referring to the accuracy between the errors and no error, the success rates of predicting a status pattern fluctuated between 80\%-90\%, while the success rates of predicting a normal state fluctuated between 75\%-90\%. The last percentages were usually lower than first ones when analyzing the confusion matrices. It is shown that there were more false positives than false negatives. That is, there were more erroneous predictions in those cases where a status pattern was predicted when in fact, the system was running well, than in those cases where a normal state was predicted when actually, there was a status pattern. In the particular context of this application, the possible consequences of an undetected fault are much greater than when detecting a fault erroneously. Therefore, although the accuracy of no errors was low, it would not have such a great consequence.
	
	The accuracy of the predictive models that were individualized by classes fluctuated too much from one class to another, so it was difficult to group them. Nevertheless, it has been shown that the higher the number of classes, the greater will be the differences between each other. While for the majority classes the success rate was 80\%-96\%, which is similar to the adopted method's accuracy, for the minority classes the percentages may have decreased until 5\%. This means that although a status pattern could be satisfactorily predicted, the type of such a status pattern could be badly predicted, especially if it was of a minority class.
	
	Placing the focus on the mean values in terms of global accuracy, sensitivity and specificity of all the predictive models, the results between this approach and the one taken as reference can be compared. As it can be appreciated in Table \ref{tab:comparison}, this approach has a better global accuracy by almost a 6\%. Moreover, it has a much higher sensitivity, concretely 15\%, and thus, it predicts with a major precision a real failure. However, it has less specificity, around a 15\%. It means it has more false positives, according to what is mentioned previously.
 	
	\begin{table}[h]
		\centering
		\caption{Our proposal's predictive models vs based work's ones}
		\begin{tabu} to .47\textwidth{X[c] | X[c] | X[c] | X[c]}
			\toprule
			Global measure types & Accuracy & Sensitivity & Specificity \\
			\midrule
			Base work \cite{kusiak_prediction_2011} & 76.50\% & 77.60\% & 75.70\% \\
			Our approach & 82.04\% & 92.34\% & 60.58\% \\
			\bottomrule
		\end{tabu}%
		\label{tab:comparison}%
	\end{table}%
	
	Although this approach was developed using a more suitable technology for a faster and scalable data analysis than the adopted one, we could not compare this approach against it in terms of scalability nor required execution time, since there was not such data available.

\section{Conclusions and future work} \label{section-conclusions}
This work has presented a Big Data analytics approach for the renewable energy field. The proposed application has adapted a predictive maintenance for wind turbines by using, at first, a Big Data processing framework to generate data-driven predictive models that are based upon historical data previously stored in the cloud. Secondly, an online fault-tolerant monitoring agent, developed by a Big Data stream processing framework, predicts the state of the wind turbines every 10 minutes. Lastly, a front-end where the status of the wind turbines can be visualized in real-time has been developed.

The experimentation has shown that the application has obtained an optimal overall success. However, the accuracy of the predictive models for predicting a concrete status pattern has margins for improvement. This has been due in a grand part to an unbalanced dataset for generating the predictive models.

This application would be improved by balancing the dataset before generating the predictive models. It would also be prudent to do some scalability tests in the cloud to see how much it could scale. Finally, an online learning could be performed to maintain the predictive models always updated, and thus, have the models adapted to the actual operating state of the wind turbines.

\bibliographystyle{IEEEtran}
\bibliography{IEEEabrv,paperTFM}

\end{document}